\documentclass[12pt]{article}
\newcommand{\fr}{ \frac}

\newcommand{\pr}{\prime}

\begin{document}
 \begin{center}
 {\Large\bf   Casimir energy inside a triangle }
 \end{center}
 \vspace{5mm}
 \begin{center}
 H. Ahmedov  and I. H. Duru
 \end{center}

 \noindent
 1.Feza G\"ursey Institute,  P.O. Box 6, 81220,
 \c{C}engelk\"{o}y, Istanbul, Turkey \footnote{E--mail :
  hagi@gursey.gov.tr and duru@gursey.gov.tr}.

 \vspace{5mm}
 \begin{center}
 {\bf Abstract}
 \end{center}
 For certain class of triangles ( with angles proportional to
$\fr{\pi}{N}$, $N\geq 3$ ) we formulate image method by making use of the
group $G_N$ generated by reflections with respect to the three lines which
form the triangle under consideration. We formulate the renormalization
procedure by classification of subgroups of $G_N$ and corresponding fixed
points in the triangle. We also calculate Casimir energy for such
geometries, for scalar massless fields. More detailed calculation is given
for odd $N$.

\vspace{1cm}
 \noindent
 {\bf I. Introduction.}

\noindent
There is a rather restricted class of geometries, for which we have
Casimir energies in explicite forms.
To calculate the energy momentum tensor one has to solve the boundary
problem, that is one one has to obtain eigenvalues and  eigenfunctions for
the field which is confined into the given region. The eigenvalues are
usually correspond to the roots of some special functions.  For example
for the three dimensional ball \cite{BALL} or
cylindrical regions \cite{CYL} to impose the required radial boundary
conditions; one has to deal with  Bessel functions and  with the roots of
them.

For some geometries with plane boundaries the Casimir problem is easier,
especially if we can employ the method of images. The original parallel
plate geometry, and in general rectangular prisms  \cite{CUBE} are of that
type.
Using the groups generated by the reflections with respect to the surfaces
one can construct the  Green functions.
For parallel plate  the group is isomorphic to $Z$, for the three
dimensional rectangular prism for example it is  $Z^3$. However if the
rectangularity condition is dropped,  the groups generated by reflections
becomes non-commutative which is the case for present work.

We  calculate the Casimir energy for the massless scalar field\footnote{It
is well known that the energy for electromagnetic field is twice the
result one gets for the massless scalar field.},
for a certain class of triangular regions. We restrict our attention on
the triangles whose angles proportional to $\fr{\pi}{N}$, where $N$ is a
natural number greater than 2. Namely one of the angles is  $\fr{\pi}{N}$
and another is $\fr{\pi}{2}$ for even $N$ and $\pi \fr{N-1}{2N}$ for odd
$N$. We also develop a renormalization technique by reducing the problem
of finding  divergences to the classification of points of the region and
their stability subgroups.

In Section II we investigate the  structure of the group $G_N$ generated
by the reflections with respect to the three lines which form the
triangle. This group will play central role in the construction of the
Green function satisfying the Dirichlet boundary condition, and in the
renormalization procedure. We show that this group is isomorphic to the
semi-direct product of the dihedral group $D_N$ and a finite dimensional
lattices.

Section III is  devoted to the construction of the Green function for the
massless scalar field which vanishes on the boundary of the triangle.

In Section IV we formulate the renormalization procedure for the Green
function in the triangle by classification of points in the triangle and
their stability groups. The trivial group consisting of the identity
element is the stability group for any points in the triangle. The
corresponding term in the Green function is the  the free Green function
in Minkowski space which makes  an infinite contribution to the energy
momentum tensor.
Stability group of the points on a side of the triangle is generated by
the reflection operator with respect to this side. In this way terms with
surface divergences is found. Since at the vertices of the triangle the
smoothness condition is violated we also have line divergences ( vertex of
the triangle in the plane corresponds to the line in the three dimension,
that is why we call it line divergence ). We find  stability groups of
these points and corresponding singular terms in the Green function.

In Section V  we give the general expression for the energy momentum
tensor in terms of the sum over elementary power functions. We also see
that the energy density per unit length in the direction perpendicular to
triangle under consideration can be represented as the integral over the
boundary of this triangle and the integrand is the elementary power
functions. We presented the calculation for odd $N$ in detail.

 \vspace{1cm}
 \noindent
 {\bf II. Reflections in a Class of Triangles.}

 \noindent
 For $N=3, 4, 5,\dots$ and $k=1,  2, \dots N-2$  consider the triangles
$\bigtriangleup^N_k$ in $x^1x^2$-plane
 formed by the  lines
 \begin{eqnarray}
 L_1 & = & \{\overrightarrow{x}\in R^2: x^2=0 \} \\
 L_2 & = & \{ \overrightarrow{x}\in R^2: x^2=x^1\tan\upsilon \} \\
 L_3 & = & \{ \overrightarrow{x}\in R^2: x^2=(b-x^1)\tan (k\upsilon ) \}
 \end{eqnarray}
 where  $b$ is the length of the side laying on the line $L_1$  and
$\upsilon= \fr{\pi}{N}$ is the angle between $L_1$  and $L_2$.

 The actions of the reflections $Q_j$ with respect to the lines $L_j$,
$j=1,2,3$  on the vector
 \begin{equation}
 \overrightarrow{x}= \left ( \begin{array}{c}
 x^1\\
 x^2
 \end{array} \right )
\end{equation}
are given by
 \begin{equation}\label{reflections}
 Q_1 \overrightarrow{x} = p \overrightarrow{x}, \ \ \ Q_2
\overrightarrow{x} = rp \overrightarrow{x},
\ \ \ Q_3 \overrightarrow{x}= pr^k \overrightarrow{x} +
\overrightarrow{x_0},
 \end{equation}
 where
  \begin{equation}\label{BASEVECTOR}
 r=\left (
 \begin{array}{cc}
 \cos 2\upsilon & - \sin 2\upsilon \\
 \sin 2\upsilon & cos 2\upsilon
 \end{array}
 \right ), \ \ \
 p=\left (
 \begin{array}{cc}
 1 & 0 \\
 0 & -1
 \end{array}
 \right ), \ \ \
\overrightarrow{x_0}=(1- pr^k) \left ( \begin{array}{c}
 1 \\
 0
 \end{array}
 \right ).
 \end{equation}
 Denote by $G_N$ the group generated by the these reflections. $G_N$ is
the free group with relations. Relations between the elements $Q_1$, $Q_2$
and $Q_3$ can be obtained from the formulas (\ref{reflections}),
and from the properties of the rotation  $r$ and  reflection $p$ operators
 \begin{equation}\label{periodic}
 r^N=1, \ \ \ p^2=1, \ \ \  pr=r^{N-1}p, \ \ \  r^k
\overrightarrow{x_0}=-p\overrightarrow{x_0}.
 \end{equation}
 Some of the obvious relations are
 \begin{equation}
 Q_j^2  =  1,  \ \ \   (Q_1Q_2)^N  = 1,
\end{equation}
 from  which we conclude that the reflections $Q_1$ and $Q_2$ generate the
finite subgroup
 \begin{equation}
 D_N=\{ r^s,  pr^s, s=0,1,\dots N-1 \}
 \end{equation}
 which is the dihedral group of dimension $2N$.
 Consider the linear space $V_N$ which consists of the vectors
 \begin{equation}
 \overrightarrow{\xi}=\sum_{s=0}^{N-1} n_s \overrightarrow{x}_s,
 \end{equation}
 where $n_s$ are integers and
 \begin{equation}\label{vector}
 \overrightarrow{x}_s = r^s\overrightarrow{x}_0.
 \end{equation}
The equalities
\begin{eqnarray}
r\overrightarrow{x}_s = \overrightarrow{x}_{s+1}, \ \
p\overrightarrow{x}_s = \overrightarrow{x}_{N-s+k}
\end{eqnarray}
imply that  $D_N$ is the  automorphism group of the linear space $V_N$.
The action of $D_N$ is given in the natural way
\begin{equation}\label{representation}
\pi (q) \overrightarrow{\xi} = q\overrightarrow{\xi}; \ \ \  q\in D_N.
\end{equation}
Since $V_N$ is a vector space over the integer numbers, unlike the spaces
over the real numbers, the
dimension $\mid V_N\mid $ is not necesserely equal to the dimension of the
vectors $\overrightarrow{x}_s$. It may be larger, that is in our case may
be greater than two.  For example  the dimensions of $V_5$, $V_8$ are
four; while the dimensions of   $V_3$, $V_6$ and $V_4$ are
two\footnote{the vector spaces $V_3\equiv V_6$ and $V_4$ are known as the
hexagonal and square lattices \cite{ARM}}. A detailed discussion of this
problem is given in the Appendix.

The group $G_N$ is the subgroup of the semidirect product group $D_N\star
V_N$. In fact for any element $g\in G_N$ one can find the pair of elements
$q\in D_N$ and $\overrightarrow{\xi}\in V_N$ as
\begin{equation}\label{repr}
g\overrightarrow{x}=q\overrightarrow{x}+\overrightarrow{\xi}\equiv
(q,\overrightarrow{\xi}) \overrightarrow{x}; \ \ \
\overrightarrow{x}\in R^2.
\end{equation}
In particular
\begin{equation}\label{semidirect}
Q_1=(p,0), \ \ \ \ Q_2=(rp,0), \ \ \  Q_3=(pr^k, \overrightarrow{x}_0).
\end{equation}
$G_N$ contains two subgroups:  $D_N$ and the one generated by $Q_3$. Since
$V_N$ does not contain invariant subspaces with respect to
(\ref{representation}) we conclude that there is no subgroup in the
semidirect product group  which contains $D_N$ and the group generated by
$Q_3$ simultaneously. This fact implies that  $G_N$ is isomorphic to
$D_N\star V_N$. In the special case of $\mid V_N\mid =2$ this group is
called the wallpaper group \cite{ARM}.

\vspace{1cm}
\noindent
{\bf III. Construction of the Green Function in the Triangles without
obtuse Angles}

 \vspace{2mm}
 \noindent
Consider the  representation of the group $G_N$ in the space of functions
on the four dimensional Minkowski space
 \begin{equation}
 T(g) f(x)=f(g x).
 \end{equation}
Here the action of the group $G_N$ is given by substitution
$\overrightarrow{x}\rightarrow x$, $\overrightarrow{\xi}\rightarrow \xi$,
$p\rightarrow P$, $r\rightarrow R$  where
\begin{eqnarray}
R =
\left ( \begin{array}{ccc}
1 & 0 & 0  \\
0 & r & 0 \\
0 & 0 & 1
\end{array}
\right ), \ \
P =
\left ( \begin{array}{ccc}
1 & 0 & 0  \\
0 & p & 0 \\
0 & 0 & 1
\end{array}
\right ),
\end{eqnarray}
which are  $4\times 4$ matrices and
\begin{eqnarray}
\xi = \left ( \begin{array}{cc}
0 \\
\overrightarrow{\xi} \\
0
\end{array}
\right ), \ \
x = \left ( \begin{array}{cc}
x^0 \\
\overrightarrow{x} \\
x^3
\end{array}
\right )
\end{eqnarray}
which are  $4$ dimensional column vectors.

Using (\ref{semidirect}) one can verify  that the operator
 \begin{equation}\label{operator}
 \textbf{O}=\sum_{n\in Z} \sum_{s=0}^{N-1} ( T((R^s,\xi))-T((PR^s,\xi)))
 \end{equation}
 satisfies the following  property
 \begin{equation}
 T(Q_j) \textbf{O} = - \textbf{O}.
 \end{equation}
In (\ref{operator}) $n=(n_0, n_1,\dots, n_{\mid V_N\mid -1})$ is
multyindex and
\begin{equation}
\xi =\sum_{t=0}^{\mid V_N\mid  -1} n_t x_t
\end{equation}
where
\begin{equation}
x_s = \left ( \begin{array}{cc}
x^0 \\
\overrightarrow{x}_s \\
x^3
\end{array}
\right )
\end{equation}
and $\overrightarrow{x}_s$ are the base vectors described in the previous
section.

It is obvious that if we define a function  $\textbf{O} f(x)$, it must
vanish  on the lines $L_j$ of reflections $Q_j$; the  fact that we make
use in the construction of the Green function inside the triangle
$\bigtriangleup^N_k$,  satisfying the Dirichlet boundary conditions. Since
the operator $\textbf{O}$ commutes with the Klein- Gordon
operator ( which is invariant under translations, rotations and
reflections ) the function
\begin{equation}\label{green}
 K (x,x^\pr) \equiv \textbf{O} G(x,x^\pr)= \sum_{n\in Z} \sum_{s=0}^{N-1}
( G (R^s x+\xi, x^\pr)-
G (PR^s x+\xi ,x^\pr)),
\end{equation}
satisfies the equation
\begin{equation}\label{delta}
\eta^{\mu\nu}\fr{\partial^2}{\partial x^\mu\partial x^\nu} K (x, x^\pr) =
\textbf{O}\delta(x-x^\pr)
\end{equation}
for any $x, x^\pr\in R^2\times\Delta^N_k$, and the boundary condition
\begin{equation}
K(x,x^\pr) \mid_{x\in R^2\times \partial\Delta^N_k} =0,
\end{equation}
where $\partial\Delta^N_k$ is the boundary of the triangle $\Delta^N_k$.
$G$ is
is the Green function in the Minkowski space with the metric $\eta =
(1,-1,-1,-1)$
\begin{equation}
G(x,x^\pr)=-\fr{1}{4\pi^2}\fr{1}{\mid x-x^\pr\mid^2}.
\end{equation}

The function $K (x, x^\pr )$ is  the Green function if the right hand side
of (\ref{delta}) is  delta function
\begin{equation}
 \textbf{O} \delta (x-x^\pr) = \delta (x-x^\pr)
\end{equation}
 for any $x, x^\pr\in R^2\times\Delta^N_k$. The above condition implies
that
for any $(q, \overrightarrow{\xi})\neq (1, 0)$ and for any
$\overrightarrow{x}, \overrightarrow{x}^\pr\in\Delta^N_k$
\begin{equation}
\delta (q\overrightarrow{x}+\overrightarrow{\xi}-\overrightarrow{x}^\pr) =
0
\end{equation}
must be satisfied.
 In other words any points inside the triangle should be the
representative of different orbits of the coset space $R^2/G_N$. The
orbits of the coset space $R^2/D_N    $
are
\begin{equation}
[\overrightarrow{x}]=\{ r^s \overrightarrow{x}, pr^s\overrightarrow{x}: \
\  s=0,\dots , N-1\}.
\end{equation}
It is clear that we can identify this coset space with  region $X$ between
two lines $L_1$ and $L_2$ including the boundaries.For any orbit in
$R^2/D_N$ there exist a unique representative in $X$. Since the group
$G_N$ is generated by the elements of $D_N$ and $Q_3$ the problem of
constructing the coset space $R^2/G_N$ reduces to finding the subspaces
$Y$ of $X$ such that the reflection $Q_3$  maps $Y$ into $X$. Consider the
area between three lines $L_j$, which is the triangle under consideration.
The previous condition implies that the two angles $k\upsilon$ and
$s\upsilon$ of the triangle between the lines $L_1$, $L_3$ and $L_2$,
$L_3$ must be less than or equal to $\fr{\pi}{2}$. The restrictions
\begin{equation}\label{equation}
k \upsilon \leq \fr{\pi}{2}, \ \ \  s\upsilon\equiv \pi -
(k+1)\upsilon\leq \fr{\pi}{2}
\end{equation}
with solutions
\begin{equation}\label{k}
k= \{
\begin{array}{cc}
\fr{N}{2}, \   for \ even \ N \\
\fr{N-1}{2},  \ for \ odd \ N
\end{array}
\end{equation}
imply that for triangles without obtuse angle the function $K(x, x^\pr)$
in (\ref{green}) is indeed the Green function. Note that the equations
(\ref{equation}) have also been solved by  $k=\fr{N-2}{2}$ for even $N$.
In this  case $s=\fr{N}{2}$. For $k=\fr{N}{2}$ we have  $s=\fr{N-2}{2}$.
Therefore this solution   is  congruent to the previous one; that is
$\Delta^N_{\fr{N}{2}}$ goes to $\Delta^N_{\fr{N-2}{2}}$ when the length
$b$ goes to $b\cos\upsilon$.

Finally we like to remark that, for massive fields, instead of $G(x,
x^\pr)$ one has to put  the Green function for
the massive scalar fields in the Minkowski space in ({\ref{green}).

\newpage
%\vspace{1cm}
\noindent
{\bf IV. Renormalization of the Green Function }

 \vspace{2mm}
 \noindent
In  polygonal regions there are three types of singular terms that to be
subtracted to obtain renormalized Green function: Free space term, surface
and vertex terms.

Inspecting (\ref{green}) we observe that the term
\begin{equation}
T(g) G(x, x^\pr )= G(gx, x^\pr )
\end{equation}
leads singularity whenever $gx=x$; that is, the  singularities arise at
the elements of the group $G_N$ which leave the points fixed.  The
renormalization problem is then reduced to the classification of the
points of  the region and
their stability subgroups.

The identity element ( which is the trivial subgroup ) leaves all points
fixed. The  term $T((1, 0))G(x, x^\pr )$ in the  (\ref{green}) therefore
gives the volume singularity and is  the free Green function.

The points on the line $L_j$ are left fixed by the reflection $Q_j$. The
group generated by $Q_j$ is then the  stability subgroup for the line
$L_j$. Since the identity element of the two-dimensional reflection group
is already employed in the volume renormalization, the surface singularity
terms in (\ref{green}) are
\begin{equation}
K_S (x, x^\pr ) = \sum_{j=1}^3 T(Q_j) G(x, x^\pr ).
\end{equation}

To discuss the vertex singularities, let us first consider the vertex at
the intersection point of the
lines $L_1$ and $L_2$. The $N$ dimensional subgroup  generated by the
element $Q_1Q_2$  is the stability subgroup of this vertex. The divergence
term at the vertex we consider is
\begin{equation}
K_{L_1L_2} (x, x^\pr ) = \sum_{j=1}^{N-1} T((Q_1Q_2)^j) G(x, x^\pr ).
\end{equation}
The element  $Q_1Q_3$ generates the stability subgroup of the vertex  at
the intersection point of the lines
$L_1$ and $L_3$. Due to restriction (\ref{k})  and  $Q_1Q_3=(r^k,
-r^k\overrightarrow{x}_0)$ we conclude that the dimension of this group is
$2$ for even $N$ and $N$ for odd $N$. Therefore we have
\begin{equation}
K_{L_1L_3} (x, x^\pr ) = \sum_{j=1}^{L-1} T((Q_1Q_3)^j) G(x, x^\pr ),
\end{equation}
where $L$ is the dimension of the stability group, that is $L=2$ if  $N$
is even and $L=N$ if $N$ is odd.

Finally let us consider the third vertex which is  the  intersection point
of the lines $L_2$ and $L_3$. The stability group of this point is
generated by the element $Q_2Q_3$. One can verify that the dimension $D$
of this group is
\begin{equation}
D= \{
\begin{array}{cc}
N \   for \ even \ \fr{N}{2} \\
\fr{N}{2}  \ for \ odd \  \fr{N}{2} \\
N \ for \ odd N
\end{array}
\end{equation}
and the corresponding singular line terms are
\begin{equation}
K_{L_2L_3} (x, x^\pr ) = \sum_{j=1}^{D-1} T((Q_2Q_3)^j) G(x, x^\pr ).
\end{equation}
Collecting all the above terms we arrive at
\begin{equation}
K_L (x, x^\pr ) = \sum_{j=1}^{N-1}( T((Q_1Q_2)^j)+ T((Q_1Q_3)^j) +
T((Q_2Q_3)^j) ) G(x, x^\pr ).
\end{equation}
for odd $N$; and,
\begin{equation}
K_L (x, x^\pr ) = (T(Q_1Q_3)+ \sum_{j=1}^{N-1}( T((Q_1Q_2)^j)+
T((Q_2Q_3)^j) ) ) G(x, x^\pr ).
\end{equation}
for even  $\fr{N}{2}$;  and
\begin{equation}
K_L (x, x^\pr ) = (T(Q_1Q_3)+ \sum_{j=1}^{N-1} T((Q_1Q_2)^j)+
\sum_{j=1}^{\fr{N}{2}-1} T((Q_2Q_3)^j)  ) G(x, x^\pr ).
\end{equation}
for odd  $\fr{N}{2}$. Subtracting all divergences  from  (\ref{green}) we
obtain the  renormalized Green function
\begin{equation}\label{renorm}
K_r(x, x^\pr) = K(x, x^\pr)-G(x, x^\pr)-K_S(x, x^\pr)-K_L(x, x^\pr).
\end{equation}
Before closing this section we like to emphasize that if the method of
images is applicable to a geometry, the stability group classification is
quite reliable approach to the renormalization.

\vspace{1cm}
\noindent
 {\bf V. Energy Momentum Tensor}

\vspace{2mm}
\noindent
The energy momentum tensor for conformally coupled massless scalar field
is given by \cite{CER}
\begin{equation}
T_{\mu\nu}=\fr{2}{3}\partial_\mu\phi\partial_\nu\phi-\fr{1}{6}\eta_{\mu\nu}\partial_\rho\phi\partial^\rho\phi-
\fr{1}{3}\phi\partial_\mu\partial_\nu\phi+\fr{1}{12}\eta_{\mu\nu}\phi\partial_\rho\partial^\rho\phi.
\end{equation}
Since the vacuum expectation value of the product of two scalar fields is
the Green function, we can express
the energy momentum tensor in the  region we study as
\begin{equation}
T_{\mu\nu}= \lim_{x\rightarrow x^\pr}
[\fr{1}{3}(\partial_\mu\partial^\pr_\nu +\partial^\pr_\mu\partial_\nu)
-\fr{\eta_{\mu\nu}}{6} \eta^{\sigma\rho} \partial_\sigma\partial^{\pr\rho}
- \fr{1}{6}(\partial_\mu\partial_\nu +\partial^\pr_\mu\partial^\pr_\nu)]
K_r(x,x^\pr)
\end{equation}
where $\partial_\mu\equiv\fr{\partial}{\partial x^\mu}$, and
$K_r(x,x^\pr)$ is given by (\ref{renorm}).

The vacuum energy density  in particular is given by the following
expression
\begin{equation}\label{CASIMIR}
T_{00} = \fr{1}{6\pi^2 } \sum_{(n, s)}
[ \fr{\mid (pr^s+1)\overrightarrow{\xi}\mid^2}{\mid
(pr^s-1)\overrightarrow{x}+\overrightarrow{\xi}\mid^6}-
\fr{2+\cos (2s\upsilon )}{\mid
(r^s-1)\overrightarrow{x}+\overrightarrow{\xi}\mid^4}].
\end{equation}
The summation runs over the indices  $n=(n_0,n_1,\dots , n_{\mid
V_N\mid-1})\in Z^{\mid V_N\mid}$,
and $s=0,1,\dots , N-1$. The terms  corresponding to the singularities
described in the previous section should be dropped in ({\ref{CASIMIR}).
The term in the second summation in ({\ref{CASIMIR}) with $(n=0, s=0)$ is
the energy of the free vacuum. The  terms with $(n=0, s=0)$, $(n=0, s=1)$
and $(n=(1,0,\dots , 0), s= k)$ in the first summation of (\ref{CASIMIR})
are  surface divergence terms. Note that  due to the  $\mid
(pr^s+1)\overrightarrow{\xi}\mid^2$ factor
they are automatically zero. This is not surprising since it is known that
the energy momentum tensor for comformally
coupled scalar field  is finite on  flat surfaces \cite{DAVIES}. Terms
$(n=0, s=1,2,\dots , N-1)$ in the second
summation of (\ref{CASIMIR}) are line divergent ones in the vertex which
is the intersection point of the lines $L_1$ and $L_2$. Using the results
of the previous section one may also find the vertex divergence terms for
other
two vertices.

Integrating the above energy density over the triangle $\Delta^N_k$ we
arrive at the energy density $E$ per unit length in $x^3$ direction. First
we observe the existence of  two exact forms
\begin{eqnarray}
\omega^s_1 & =& \fr{2+\cos (2s\upsilon
)}{2\mid(r^s-1)\overrightarrow{x}+\overrightarrow{\xi} \mid^4} dx^1\wedge
dx^2, \\
\omega^s_2 & =& \fr{\mid
(pr^s+1)\overrightarrow{\xi}\mid^2}{\mid\overrightarrow{v}(pr^s)\mid^6}dx^1\wedge
dx^2.
\end{eqnarray}
They can be rewritten as  $\omega^s_j=d\Omega^s_j $, where
\begin{eqnarray}
\Omega^s_1 & =& D_s\fr{((r^s-1)\overrightarrow{x}+\overrightarrow{\xi})^2
dx^1-((r^s-1)\overrightarrow{x}+\overrightarrow{\xi})^1 dx^2}
{\mid (r^s-1)\overrightarrow{x}+\overrightarrow{\xi} \mid^4}, \ \  s\neq 0
\\
\Omega^s_2 & =& \fr{\mid (pr^s+1)\overrightarrow{\xi}\mid^2
}{\mid(pr^s-1)\overrightarrow{x}+\overrightarrow{\xi}\mid^6}
(r^{\fr{s}{2}}\overrightarrow{x})^1 d(r^{\fr{s}{2}}\overrightarrow{x})^2,
\end{eqnarray}
where $D_s=\fr{2+\cos (2s\upsilon )}{6(1-\cos (2s\upsilon ))}$. By making
use of  the Stokes theorem one
can convert the integration over the triangle $\Delta^N_k$ into the
integral over its boundary $\partial\Delta^N_k$.
The energy density per unit length in $x^3$ direction is then
\begin{equation}\label{VACUUM}
E=-\fr{1}{6\pi^2 } \langle  J_0 + J_1 - J_2 \rangle
\end{equation}
where $S^N_k$ is the area of the triangle $\Delta^N_k$ and
\begin{eqnarray}
J_0 & = & \sum_{n\in Z^{\mid V_N\mid}} \fr{3S^N_k}{\mid
\overrightarrow{\xi}\mid^4} \\
J_1 & = &   \sum_{n\in Z^{\mid V_N\mid}}\sum_{s=1}^{N-1}
\int_{\partial\Delta^N_k} \Omega_s^1 \\
J_2 & = &   \sum_{n\in Z^{\mid V_N\mid}}\sum_{s=0}^{N-1}
\int_{\partial\Delta^N_k} \Omega_s^2
\end{eqnarray}
In $J_j$ we take summation over all values of $n$ and $s$. The brackets
$\langle \rangle$ in (\ref{VACUUM}) means
that we have to drop the singularity  terms. $J_j$ may be divergent  if
$\mid V_N\mid$ is greater than three. However their  difference  should be
finite. We have to  treat each term as formal series. Then collecting them
together we obtain the final result.

Let us restrict our attention to the case of odd $N$, for which the vector
space $V_N$ appears to be invariant under the half angle rotations
$r^{\fr{s}{2}}$. This can be shown from the identity
\begin{equation}
r^{\fr{N}{2}}\overrightarrow{\xi}=-\overrightarrow{\xi}.
\end{equation}
Using the relations
\begin{equation}
(pr^s\pm 1)= r^{-\fr{s}{2}} (p\pm 1 ) r^{\fr{s}{2}}, \ \ \ (r^s-1)=-a_s
r^{\fr{s}{2}+\fr{N}{4}}
\end{equation}
with $a_s=2\sin (s\upsilon )$ and making the reparametrization in the
multyindex $n$ which is equivalent  to the
change of variable $r^{\fr{s}{2}}\overrightarrow{\xi}\rightarrow
\overrightarrow{\xi}$  we arrive at
\begin{equation}\label{EQ1}
J_1=- \sum_{n\in Z^{\mid V_N\mid}}\sum_{s=1}^{\fr{N-1}{2}}D_s
\int_{\partial\Delta_{\fr{N}{2}}}
\fr{(a_s\overrightarrow{z}-\overrightarrow{\xi})^2 dz^1
-(a_s\overrightarrow{z}-\overrightarrow{\xi})^1 dz^2}{\mid
a_s\overrightarrow{z}-\overrightarrow{\xi}\mid^4}
\end{equation}
\begin{equation}
J_2= \sum_{n\in Z^{\mid V_N\mid }}\sum_{s=0}^{N-1} \int_{\partial\Delta_s}
\fr{\mid (p+1)\overrightarrow{\xi}\mid^2 }{\mid
(p-1)\overrightarrow{y}+\overrightarrow{\xi} \mid^6} y^1dy^2.
\end{equation}
Here  we  have used the short notation  $\Delta_0\equiv \Delta^N_k$ and
$\Delta_s= r^{\fr{s}{2}} \Delta_0$,  that is the triangle $\Delta_s$ is
$\Delta_0$ rotated by the angle $s\upsilon$.                We also used
the symmetry $s\rightarrow N-s$ in (\ref{EQ1}) to reduce  the summation
over $s$. Denote by $a_0$, $a_1$ and $c_0$ the  sides of the triangle
$\Delta_0$ laying on the lines $L_1$, $L_2$ and $L_3$.
Then $a_s$, $a_{s+1}$ and $c_s$ will be the sides of the triangle
$\Delta_s$, that is $a_{s+1}= r^{\fr{s}{2}} a_s$
and $c_{s+1}= r^{\fr{s}{2}} c_s$. Since the orientation on the side
$a_{s+1}$ of the triangle $\Delta_s$ is opposite to the one on the side
$a_{s+1}$ of $\Delta_{s+1}$ we have
\begin{equation}
J_2= \sum_{n\in Z^{\mid V_N\mid }}\int_{U} \fr{\mid
(p+1)\overrightarrow{\xi}\mid^2 }{\mid
(p-1)\overrightarrow{y}+\overrightarrow{\xi} \mid^6} y^1dy^2,
\end{equation}
where the integration contour $U=a_0\cup b_0\cup b_1\cup\cdots b_{N-1}\cup
a_N$ oriented  anti clock wise.
On the sides $a_0$, $a_N$ and $c_{\fr{N-1}{2}}$ we have $y^2=const$, that
is these sides do not make contribution in $J_2$. We also  observe that
reflection operator $-p$ with respect to the $y^2$ axis send   $c_j$ to
$c_{N-j-1}$ with opposite orientation. Since the one form in the  integral
change sign under reflection $\overrightarrow{y}\rightarrow -p
\overrightarrow{y}$ one can rewrite the above expression as
\begin{equation}
J_2 = 2 \sum_{n\in Z^{\mid V_N\mid }}\sum_{j=0}^{\fr{N-3}{2}} \int_{c_j}
\fr{\mid (p+1)\overrightarrow{\xi}\mid^2 }{\mid
(p-1)\overrightarrow{y}+\overrightarrow{\xi} \mid^6} y^1dy^2.
\end{equation}
For given $c_j$ we construct closed contour in the following way. From the
end points of $c_j$ draw lines which are parallel to the $y^1$-axis. They
intersect $y^2$ axis at the points $b\sin j\upsilon$ and $b\sin
(j+1)\upsilon$.
The interval between these two points, $c_j$ and two intervals between
them, which  are parallel to $y^1$-axes form the desired contour which we
denote by $C_j$. Since at $y^1=0$ and $y^2=const$ the one form in the
above integral is zero. We then  have
\begin{equation}
J_2 = 2 \sum_{n\in Z^{\mid V_N\mid }}\sum_{j=0}^{\fr{N-3}{2}} \int_{C_j}
\fr{\mid (p+1)\overrightarrow{\xi}\mid^2 }{\mid
(p-1)\overrightarrow{y}+\overrightarrow{\xi} \mid^6} ( y^1-A_j y^2
-B_j)dy^2
\end{equation}
where we have added the exact forms, which are the second and third terms
in the bracket, for
integral of the exact form over the closed form is zero. We choose the
coefficients $A_j$ and $B_j$ to satisfy
\begin{equation}
y^1-A_j y^2 -B_j=0,
\end{equation}
for $\overrightarrow{y}\in c_j$, that is to make the value of the one form
zero on the side $c_j$. We have
\begin{equation}
A_j= -\fr{\cos\upsilon (k-j)}{\sin\upsilon (k-j)}, \ \ \
B_j= b \fr{\sin\upsilon k}{\sin\upsilon (k-j)}
\end{equation}
where $k=\fr{N-1}{2}$. Non zero contribution to the integral comes only
from the integration over the interval laying on the $y^2$ axis:
\begin{equation}
J_2 = 8 \sum_{n\in Z^{\mid V_N\mid }}\sum_{j=0}^{\fr{N-3}{2}} \int_{ b\sin
j\upsilon )}^{b\sin (j+1)\upsilon}
 \fr{(\xi^1)^2( A_j t +B_j)dt}{((\xi^1)^2+ (2t-\xi^2)^2)^3}
\end{equation}
or
\begin{equation}\label{J_2}
J_2 = \fr{1}{2b^2\sin^2 k\upsilon }
\sum_{n\in Z^{\mid V_N\mid }}\sum_{j=0}^{\fr{N-3}{2}} \fr{\eta_1^2}{\sin
(k-j)\upsilon}
\int_{f_j}^{f_{j+1}}
dx  \fr{ 1- x\cos ((k-j)\upsilon) }{(\eta_1^2+ (x-\eta_2)^2)^3},
\end{equation}
where $f_j=\fr{\sin j\upsilon}{\sin k\upsilon}$ and
$\overrightarrow{\xi}=2b \sin k\upsilon \overrightarrow{\eta}$ or
\begin{equation}
\overrightarrow{\eta} = \sum_{t=0}^{\mid V_N\mid -1} n_t
\overrightarrow{x}^\pr_t, \ \ \
\overrightarrow{x}^\pr_t=r^{t+\fr{1}{4}}
\left ( \begin{array}{c}
 1 \\
 0
 \end{array}
 \right ).
\end{equation}
(We also used lower indices for the vector $\overrightarrow{\eta}$ ).

Now consider the expression (\ref{EQ1}). Rotation $r^\fr{1}{2}$ maps
interval $a_{\fr{N}{2}}$ on $a_{\fr{N}{2}+1}$ which  has the opposite
orientations. Since the expression under the integral is invariant under
transformation $\overrightarrow{z}\rightarrow
r^{\fr{1}{2}}\overrightarrow{z}$, the contributions from these intervals
cancel each other. $J_1$ is nonzero on  the interval $c_{\fr{N}{2}}$. Let
us make change of variable
$\overrightarrow{y}=r^{-\fr{N+1}{4}}\overrightarrow{z}$. Then
$c_{\fr{N}{2}}$ goes to $c_{-\fr{1}{2}}$ on which
$y^1=b\cos\fr{\upsilon}{2}$ and $y^2\in [-b\sin\fr{\upsilon}{2},
b\sin\fr{\upsilon}{2} ]$. Therefore
\begin{equation}
J_1= \sum_{n\in Z^{\mid V_N\mid}}\sum_{s=1}^{\fr{N-1}{2}}D_s
\int_{-b\sin\fr{\upsilon}{2}}^{b\sin\fr{\upsilon}{2}}
\fr{(ba_s\cos\fr{\upsilon}{2} - \xi^1) dy}
{((ba_s\cos\fr{\upsilon}{2} - \xi^1)^2+(a_s y -\xi^2)^2)^2}
\end{equation}
or
\begin{equation}
J_1=\fr{\sin\fr{\upsilon}{2}}{8b^2\sin^3 k\upsilon}
\langle  \sum_{n\in Z^{\mid V_N\mid}}
\sum_{s=1}^{\fr{N-3}{2}}\fr{D_s}{f_s}
\int_{-f_s}^{f_s}
\fr{(f_s\cos\fr{\upsilon}{2} - \eta_1) dx}
{((f_s\cos\fr{\upsilon}{2} - \eta_1)^2+(x\sin\fr{\upsilon}{2}
-\eta_2)^2)^2}.
\end{equation}
Note that in the above expression we have dropped the term
$s=\fr{N-1}{2}$. Since  $f_{\fr{N-1}{2}}=1$ by the  reparametrization
$\overrightarrow{\eta}\rightarrow\overrightarrow{\eta}+\overrightarrow{x}_0^\pr$
we can rewrite it as
\begin{equation}
-\fr{1}{8b^2\sin^3k\upsilon}D_{\fr{N-1}{2}}
 \sum_{n\in Z^{\mid V_N\mid}}
\int_{-1}^1
\fr{\eta_1 dx}
{ \eta_1^2+((x-1)\sin\fr{\upsilon}{2} -\eta_2)^2)^2}
\end{equation}
which is odd function in $\eta_1$ variable. Therefore it is zero.

Finally let us consider the special case when $N=3$ and $k=1$. We have
$\upsilon =\fr{\pi}{3}$  and $\overrightarrow{\xi}=
\sqrt{3} b\overrightarrow{\eta}$ with
\begin{equation}
\overrightarrow{\eta} =  \fr{n_0}{2} \left ( \begin{array}{c}
 \sqrt{3} \\
 1
 \end{array}
 \right ) + \fr{n_1}{2}
\left ( \begin{array}{c}
 -\sqrt{3} \\
 1
 \end{array}
 \right ),
\end{equation}
that is $\mid V_3\mid=2$. We have $S^3_1= \fr{\sqrt{3}}{4}b^2$,  $J_1=0$
and
\begin{equation}
J_0= \fr{\sqrt{3}}{12 b^2}\sum_{n\in Z^2} \fr{1}{\mid
\overrightarrow{\eta}\mid^4},
\end{equation}
\begin{equation}
J_2 = \fr{2}{3\sqrt{3}b^2}
\sum_{n\in Z^2} \eta_1^2\int_0^1
dx  \fr{ 2- x}{(\eta_1^2+ (x-\eta_2)^2)^3}.
\end{equation}
The expressions
\begin{eqnarray}
\sum_{n\in Z^2} \eta_1^2\int_0^1
dx  \fr{ x-\eta_2}{(\eta_1^2+ (x-\eta_2)^2)^3} & = & -\fr{1}{4}\sum_{n\in
Z^2}  \fr{1}{(\eta_1^2+ (x-\eta_2)^2)^2}\mid_0^1 \\
\sum_{n\in Z^2} \eta_1^2\int_0^1
dx  \fr{1}{(\eta_1^2+ (x-\eta_2)^2)^3} & = & \sum_{n\in Z^2}  F(\eta_1,
x-\eta_2)\mid_0^1
\end{eqnarray}
due to the symmetry $\eta_1\rightarrow\eta_1$ and $\eta_2\rightarrow
\eta_2+1$ is zero.
Here
\begin{equation}
F(x, y )=\int_{-\infty}^y \fr{dz}{(z^2 +x^2)^3}.
\end{equation}
Therefore we are left with
\begin{equation}
J_2 = -\fr{2}{3\sqrt{3}b^2}
\sum_{n\in Z^2} \fr{1}{\eta_1^3}  f(\fr{\eta_2}{\eta_1}),
\end{equation}
where
\begin{equation}
f(x) =\int_{-\infty}^x \fr{dz}{(z^2 +1)^3}.
\end{equation}
The vacuum energy  density per unit length in $x^3$ direction is then
\begin{equation}
E=-\fr{2}{3\sqrt{3}\pi^2 b^2} \sum_{(l,m)\neq (0,0)} (
 \fr{1}{(3l^2+m^2)^2} + \fr{4}{9\sqrt{3} l^3} f(\fr{m}{\sqrt{3}l}) ).
\end{equation}

\vspace{1cm}
\noindent
{\bf VI Discussion  }

\vspace{2mm}
\noindent
We have calculated the Casimir energy for a class of triangles without
obtuse angle.  We applied the method of
images. Unlike the case of parallel plate or rectangular prisms, the group
generated by reflections is not abelian;
thus, the employment of the image method for triangles is not trivial.

Renormalization procedure is observed to be equivalent to the
classification of the points in the triangle and their stability
subgroups. To renormalize the Green function we subtract the terms
corresponding these stability subgroups.
Identity element is the stability subgroup for the all points, reflections
and bi-product of reflections generate the stability subgroups of points
on the planes involving the sides, and of the lines passing through the
vertices respectively.

We hope that the technique we used which essentially is based on the
employment of the groups generated by reflections from the surfaces, can
be employed for other polygonal regions. We also hope that it may even be
possible to study some other geometries with smooth boundaries, as the
limiting case of the suitable polygonal regions.  For example for an
elliptical region such a process may not be as hopeless as dealing with
the roots of Mathieu functions.

\vspace{1cm}
\noindent
{\bf Acknowledgment} Authors thank TUBA ( Turkish Academy of Sciences )
for its support.

\setcounter{equation}{0}
\def\theequation{A.\arabic{equation}}

\vspace{1cm}
\noindent
{\bf Appendix }

\vspace{2mm}
\noindent
To find the dimension $\mid V_N\mid $ of $V_N$ one has to investigate the
non zero independent solutions of the equation
\begin{equation}\label{independent}
\sum_{s=0}^{N-1} n_s \overrightarrow{x}_s=0
\end{equation}
 Assume that we have found all independent solutions
$n^t=(n^t_0,n^t_1,\dots , n^t_{N-1} )$, $t=1,\dots m$
then $\mid V_N\mid  = N-m$. From (\ref{vector}) we observe that the
equation (\ref{independent}) can be rewritten  in the  form
\begin{equation}\label{independent1}
\sum_{s=0}^{N-1} n_s r^s = 0.
\end{equation}
Let $N$ be a prime number. The periodicity condition $r^N=1$ and the
identity
\begin{equation}\label{identity}
\fr{1-x^N}{1-x} =1+x+\cdots + x^{N-1}
\end{equation}
which is valid for $x\neq 1$ imply  the solution $n^0=(1,1,\dots , 1)$  of
(\ref{independent}).
Let $N=M^l$, where $M$ is a prime  and $l$ is a  natural numbers. Then the
operators
$R_p=r^{\fr{N}{M^p}}$, $p=1,2,\dots, l$  satisfy the periodicity condition
$R_p^{M^p}=1$. This implies
$\fr{N}{M}+\fr{N}{M^2}+\cdots + \fr{N}{M^{l-1}}+1$ relations of the form
\begin{equation}
 r^{s_p} (1+R_p+\cdots + R_p^{M^p-1} ) = 0
\end{equation}
or
\begin{equation}
 r^{s_p}+r^{\fr{N}{M^p}+s_p}+\cdots + r^{\fr{N}{M^p}(M^p-1)+s_p} = 0.
\end{equation}
We denote the corresponding solutions by $n^{s_p}$, where $p=1,2, \dots ,
l$ and $s_p =0, 1, \dots \fr{N}{M^p}-1$.
The rank of the matrix  $(n^{s_p}_s)$ appears to be $\fr{N}{M}$ and we
choose solutions $n^{s_1}$ as independent set. We demonstrate it for
$N=2^3$. We have four relations
\begin{equation}\label{sol1}
 r^{s_1}+r^{4+s_1}= 0, \ \ \ s_1=0,1,2,3
\end{equation}
two relations
\begin{equation}\label{sol2}
 r^{s_2}+r^{2+s_2}+r^{4+s_2}+r^{6+s_2}= 0, \ \ \ \ s_2=0,1
\end{equation}
and one relation
\begin{equation}\label{sol3}
 1+r+ \cdot + r^7= 0.
\end{equation}
We see that the relations  (\ref{sol2}) reduce to
\begin{equation}\label{sol1}
 r^{s_2}+r^{4+s_2}= 0, \ \ \ r^{(2+s_2)}+r^{4 + (2+s_2)}= 0.
\end{equation}
Summation of four relations (\ref{sol1}) leads the relation (\ref{sol3}).
Therefore we have four independent
solutions. The general case can be proved in a similar fashion. Let now
$N=M_1^{l_1}M_2^{l_2}\cdots M_f^{l_f}$, where $M_j$
are prime numbers such that $M_1<M_2<\cdots < M_f$. In this case we have
$\fr{N}{M_1}+\fr{N}{M_2}+\cdots + \fr{N}{M_f}$ solutions $n^{s^j}$:
\begin{equation}
 r^{s^j}+r^{\fr{N}{M_j}+s^j}+\cdots + r^{\fr{N}{M_j}(M_j-1)+s^j} = 0
\end{equation}
where $j=1,2,\dots f$ and $s^j=0,\dots \fr{N}{M_j}-1$. The rank of the
matrix $A_N=(n^{s^j}_s)$ gives the number of
independent relations. The matrix $A_N$ which consist of
$\fr{N}{M_1}+\fr{N}{M_2}+\cdots + \fr{N}{M_f}$ rows and $N$
columns can be shown to have the following form
\begin{equation}
A_N=\left(
\begin{array}{cccc}
I_1 & I_1 & \cdots & I_1 \\
I_2 & I_2 & \cdots & I_2 \\
\cdots & \cdots & \cdots &  \cdots \\
I_f & I_f & \cdots & I_f \\
\end{array}
\right),
\end{equation}
where $I_j$ is $\fr{N}{M_j}\times\fr{N}{M_j}$ unit matrix. Note that the
number of $I_j$ matrices in $j^{th}$ row is $M_j$. Assume that the
relations described above  exhaust all relations of the form
(\ref{independent}). Then
we have $\mid V_N\mid  = N- rank (A_N)$. For example the vector spaces
$V_3$, $V_4$ and $V_6$ has dimension two.
We conjecture the following result
\begin{eqnarray}
\mid V_N \mid & = & N-1, \ \ \   for \ a \ prime \  N \\
\mid V_N \mid & = & N-\fr{N}{M}, \ \  for \ N=M^l \ and \ a \ prime \ M.
\end{eqnarray}

 \end{document}